\documentclass[twocolumn,aps,pra,superscriptaddress,showpacs,tightenlines]{revtex4}
\usepackage{amsmath}
\usepackage{amsfonts}
\usepackage{graphicx}
\usepackage{epsfig}
\usepackage{color}
\usepackage{hyperref}
\hypersetup{
    colorlinks=true, %set true if you want colored links
    linktoc=all,     %set to all if you want both sections and subsections linked
    linkcolor=blue,  %choose some color if you want links to stand out
    citecolor=blue}

\begin{document}

\title{Steady-state mechanical squeezing in an optomechanical system via Duffing nonlinearity}

\author{Xin-You L\"{u}}
\affiliation{School of physics, Huazhong University of Science and Technology, Wuhan 430074, China}
\affiliation{CEMS, RIKEN, Saitama 351-0198, Japan}
\author{Jie-Qiao Liao}
\affiliation{CEMS, RIKEN, Saitama 351-0198, Japan}
\author{Lin Tian}
\email{ltian@ucmerced.edu}
\affiliation{School of Natural Sciences, University of California, Merced, California 95343, USA}
\author{Franco Nori}
\email{fnori@riken.jp}
\affiliation{CEMS, RIKEN, Saitama 351-0198, Japan}
\affiliation{Department of Physics, The University of Michigan, Ann Arbor, Michigan 48109-1040, USA}

\begin{abstract}
Quantum squeezing in mechanical systems is not only a key signature of macroscopic quantum effects, but can also be utilized to advance the metrology of weak forces. Here we show that strong mechanical squeezing in the steady state can be generated in an optomechanical system with mechanical nonlinearity and red-detuned monochromatic driving on the cavity mode. The squeezing is achieved as the joint effect of nonlinearity-induced parametric amplification and cavity cooling, and is robust against thermal fluctuations of the mechanical mode. We also show that the mechanical squeezing can be detected via an ancilla cavity mode.
\end{abstract}
\maketitle

\section{Introduction~\label{sec1}} Enormous progress has been achieved in the field of cavity optomechanics in the past few years~\cite{reviews}. Examples include the preparation of mechanical modes to their quantum ground state, the demonstration of strong optomechanical coupling in the microwave and optical regimes, and the coherent state conversion between cavity and mechanical modes~\cite{Groundstate1, strongcouplingExp1, strongcouplingExp2, strongcouplingExp3,  strongcouplingExp4, strongcouplingExp5, stateconversion1, stateconversion2, stateconversion3, stateconversion4}. Given these technological advances, the effective quantum manipulation of mechanical modes becomes a promising goal.

Quantum squeezing of mechanical modes is one of the key macroscopic quantum effects that can be utilized to study the quantum-to-classical transition and to improve the precision of quantum measurements~\cite{Walls1983, squeezing_opt1, squeezing_opt2, squeezing_opt3, CavesRMP1980, CavesPRD1981, NoriPRL1996}. Thermal squeezing of mechanical modes using parametric processes and measurement-based ideas has been demonstrated in recent experiments~\cite{squeezing_exp1, squeezing_exp2, squeezing_exp3, squeezing_exp4, squeezing_exp5}. In simple schemes using parametric amplification, squeezing is limited by the so-called $3$ dB limit -- quantum noise cannot be reduced below half of the standard quantum limit -- due to the instability of the mechanical systems~\cite{MilburnWalls1981}. In recent years, a number of schemes have been proposed to generate mechanical squeezing that can go beyond the $3$ dB limit, including methods based on parametric processes, measurement- and feedback-based schemes, as well as approaches utilizing the concept of quantum reservoir engineering~\cite{parametric1, parametric2, parametric3, parametric4, parametric5, parametric6, parametric7, parametric8, parametric9, parametric10, nonlinearity1, nonlinearity2, feedback1, feedback2, feedback3, reservoir1, reservoir2, reservoir3, reservoir4}. However, quantum squeezing of mechanical modes has not been observed experimentally. Note that in recent experiments, squeezing in optical fields has been achieved in optomechanical systems~\cite{light_squeezing1, light_squeezing2, light_squeezing3}. These experiments have the potential to reach a squeezing level well below the quantum limit.
\begin{figure}
\includegraphics[width=\columnwidth,clip]{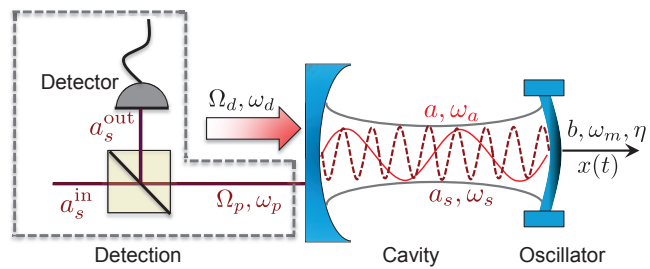}
\caption{The schematic of an optomechanical system with mechanical mode $b$ (nonlinearity $\eta$), main cavity $a$, and an ancilla cavity $a_{s}$. The pump field on cavity $a$ ($a_{s}$) is indicated by amplitude $\Omega_d$ ($\Omega_p$) and frequency $\omega_d$ ($\omega_p$). The detection circuit is enclosed by gray-dashed lines.}
\label{fig1}
\end{figure}

Here we present a method to generate strong steady-state mechanical squeezing in an optomechanical system via mechanical nonlinearity and cavity cooling. The mechanical nonlinearity required in this scheme is achieved by coupling the mechanical mode to an ancilla system, such as an external electrode or a qubit, and its magnitude far exceeds that of the intrinsic mechanical nonlinearity~\cite{nems1, nems2}. The driving on the cavity is a red-detuned monochromatic source which generates strong optomechanical coupling between the cavity and the mechanical modes and greatly reduces the thermal fluctuations of the mechanical mode. This driving, when combined with the nonlinearity of the mechanical mode, also induces a parametric-amplification process which plays a key role in generating squeezing. We find that near an optimal detuning point, strong squeezing well below the standard quantum limit can be reached even at high temperatures. Meanwhile, the red-detuned driving serves to protect the system from instability. The mechanical squeezing can be detected by homodyning the output field of an ancilla cavity mode driven by a second pump pulse. Compared with previous works, our proposal only requires one driving source on the main cavity and is robust against thermal fluctuations. The parametric-amplification process induces a huge increase in the effective mechanical frequency which strongly suppresses the quantum backaction noise. Our proposal could help the generation of strong quantum squeezing in mechanical systems.

This paper is organized as follows. In Sec.~\ref{sec2}, we introduce an optomechanical system with mechanical nonlinearity and derive its effective Hamiltonian under strong driving. In Sec.~\ref{sec3}, we study the steady-state squeezing of the mechanical mode and identify the optimal parameter regime for the squeezing. Analytical solutions of two limiting cases are presented in Sec.~\ref{sec4}, and the detection of the mechanical squeezing is discussed in Sec.~\ref{sec5}. In Sec.~\ref{sec6}, we discuss the validity of the linearization procedure and the effect of the detection on the proposed squeezing scheme. Conclusions are given in Sec.~\ref{sec7}.

\section{System~\label{sec2}} 
Consider the optomechanical system depicted in Fig.~\ref{fig1} with the Hamiltonian ($\hbar=1$)
\begin{eqnarray}
H_{t}&=&H_{c}+H_{m}-g_{0}a^{\dagger}a(b^{\dagger}+b),\label{H} \\
H_{c}&=&\delta_{a}a^{\dagger}a +\Omega_{d}(a^{\dagger}+a),\label{Hc} \\
H_{m}&=&\omega_{m}b^{\dagger}b+(\eta/2)(b^{\dagger}+b)^{4},\label{Hm}
\end{eqnarray}
where $a$ ($a^{\dagger}$) and $b$ ($b^{\dagger}$) are the annihilation (creation) operators of the cavity mode and the mechanical mode, respectively. The cavity mode (with frequency $\omega_{a}$) is described by the Hamiltonian $H_{c}$ written in the rotating frame of a monochromatic driving field with detuning $\delta _{a}$ and amplitude $\Omega_{d}$. The Hamiltonian of the mechanical mode $H_{m}$ (with frequency $\omega_{m}$) contains a Duffing nonlinear term with amplitude $\eta$. The last term in Eq.~(\ref{H}) describes the radiation-pressure interaction between the cavity and the mechanical modes with coupling strength $g_{0}$~\cite{CKLawPRA1995}. For mechanical modes in the sub-gigahertz range, the intrinsic nonlinearity is usually very weak with nonlinear amplitude smaller than $10^{-15}\,\omega_m$~\cite{nems1}. A strong nonlinearity can be produced by coupling the mechanical mode to an ancilla system~\cite{BlencowePR2004, NoriRMP2013, strong_nonlinear1, strong_nonlinear2}. For example, by coupling the mechanical mode to a qubit, a nonlinear amplitude of $\eta=10^{-4}\,\omega_m$ can be obtained (see Appendix~\ref{ssec1} for details). Other approaches can also be applied to enhance the nonlinearity, such as by softening the mechanical mode~\cite{nonlinearity1, strong_nonlinear4}. Note that nonlinearity in other forms, such as cubic potential $\eta(b+b^{\dag})^{3}$, can also be utilized to implement our scheme (see Appendix~\ref{ssec2} for details).

When including the dissipation caused by the system-bath coupling, the full dynamics of this optomechanical system is described by the master equation
\begin{equation}
\dot{\rho}=-i[H_{t},\rho]+\kappa\mathcal{D}[a]\rho+\gamma(\bar{n}_{\textrm{th}}+1)\mathcal{D}[b]\rho+\gamma\bar{n}_{\textrm{th}}\mathcal{D}[b^{\dagger}]\rho. \label{rho1}
\end{equation}
Here $\mathcal{D}[o]\rho=o\rho o^{\dagger}-(o^{\dagger}o\rho+\rho o^{\dagger}o)/2$ is the standard Lindblad superoperator for the damping of the cavity and the mechanical modes, $\kappa$ and $\gamma$ are the cavity and the mechanical damping rates, respectively, and $\bar{n}_{\rm th}$ is the thermal phonon occupation number.

\begin{figure}[ht]
\includegraphics[width=\columnwidth,clip]{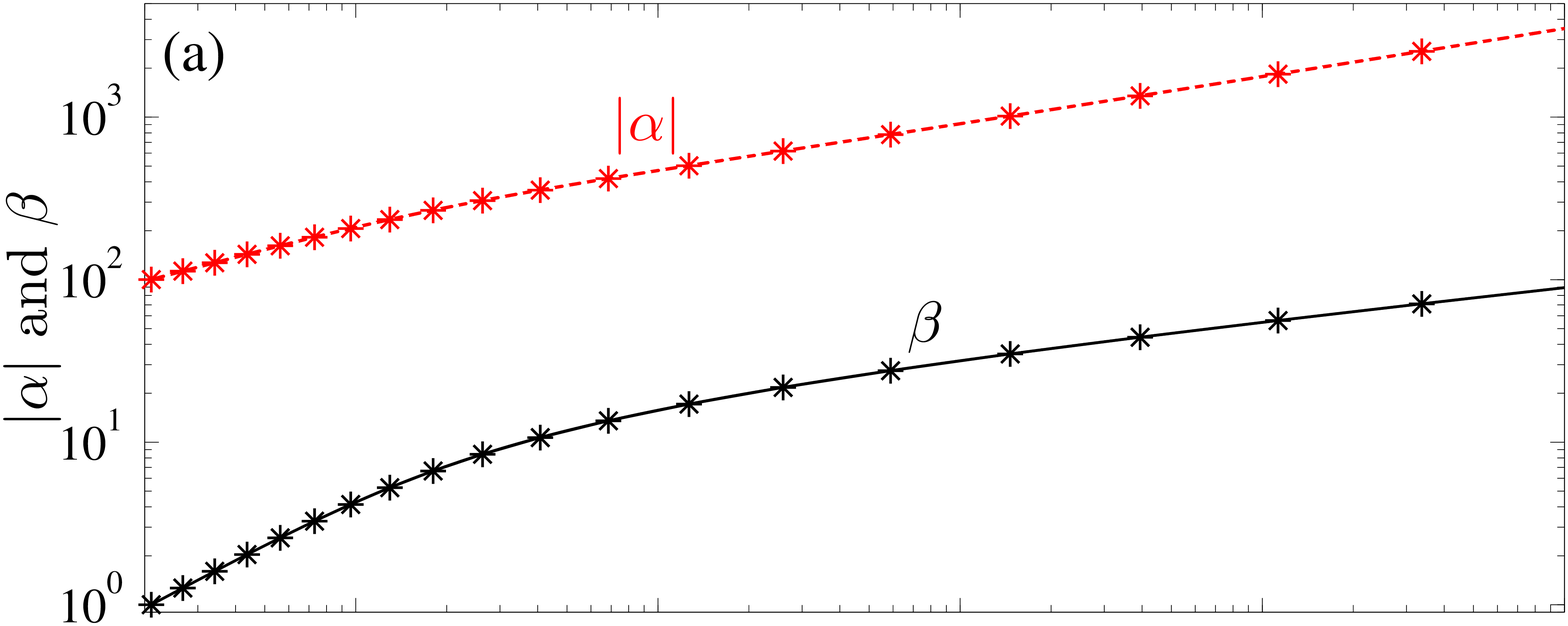}
\includegraphics[width=\columnwidth,clip]{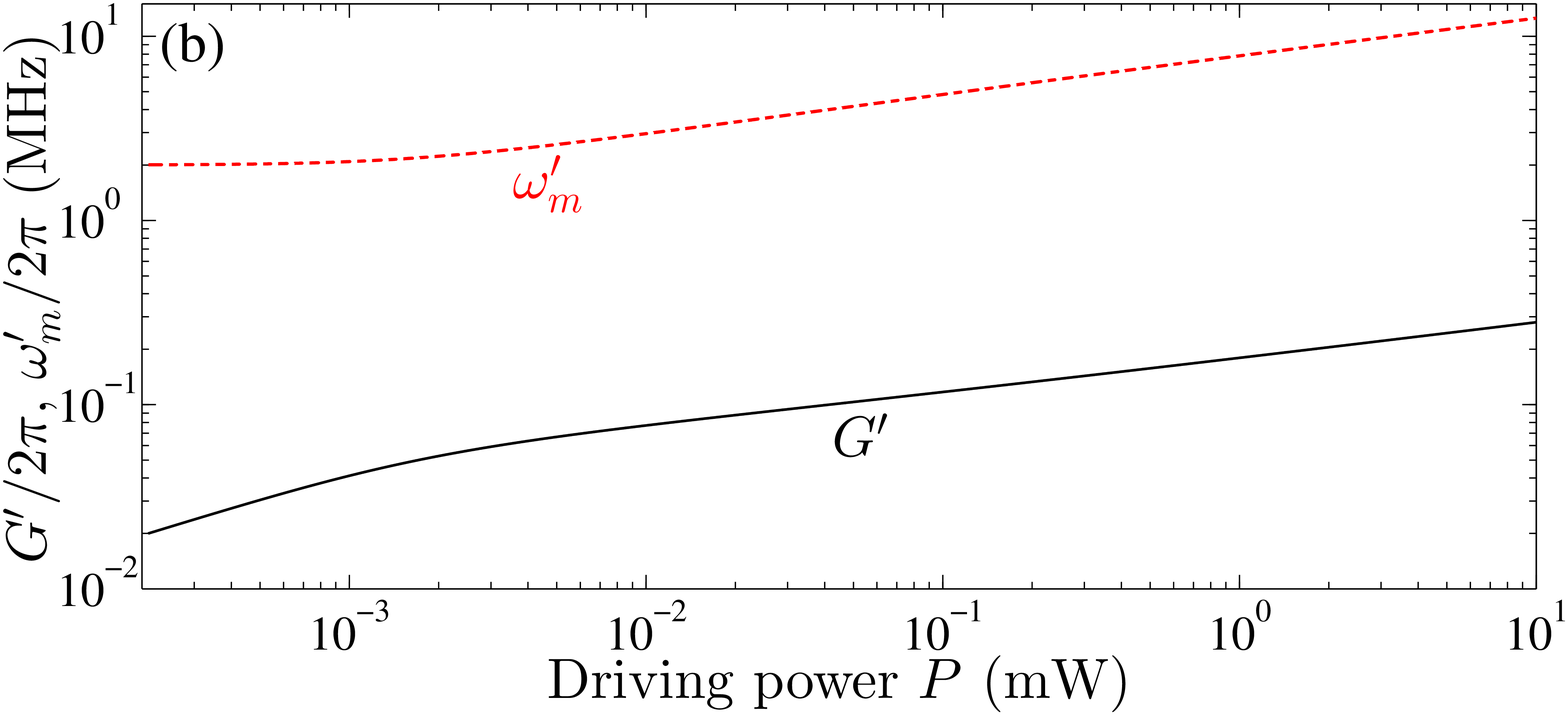}
\caption{(a) The steady-state amplitudes $|\alpha|$ and $\beta$ versus the driving power $P$. (b) The squeezed mechanical frequency $\omega_{m}^{\prime}$ and the coupling constant $G^{\prime}$ versus $P$. The asterisks in (a) are obtained with the detection circuit included (see Eq.~(\ref{detectionH})). The frequencies of the cavity modes $a$ and $a_s$ are $\omega_a/2\pi=500\,\textrm{THz}$ and $\omega_s/2\pi=1000\,\textrm{THz}$, respectively. The driving amplitudes are $\Omega_d=\sqrt{2P\kappa/\omega_{a}}$ and $\Omega_p=\sqrt{2P_s\kappa_s/\omega_{s}}$. Other parameters are $\omega_m/2\pi=2\,\textrm{MHz}$, $g_0=g_s=10^{-4}\,\omega_m$, $\eta=10^{-4}\,\omega_m$, $\kappa=\kappa_s=0.1\,\omega_m$, $\gamma=10^{-6}\,\omega_m$, and $P_s=0.1$ $\mu$W.}
\label{fig2}
\end{figure}
Strong red-detuned driving on the cavity generates large steady-state amplitudes in both the cavity and the mechanical modes. Let $\alpha$ ($\beta$) be the steady-state amplitude of the cavity (mechanical) mode under the red-detuned driving. Using the standard linearization procedure, the steady-state amplitudes can be derived by solving the following equations:
\begin{subequations}
\begin{align}
\left[-i(\delta_a-2g_0\beta)-\kappa/2\right]\alpha-i\Omega _{d}=0, \label{ss1}\\
16\eta\beta^3+(12\eta+\omega_{m})\beta-g_0|\alpha|^2=0, \label{ss2}
\end{align}
\end{subequations}
where we have dropped $\gamma$-dependent terms because $\gamma\ll \kappa, \eta$. With (moderately) strong driving on the cavity, these amplitudes satisfy $|\alpha|,\,\beta\gg1$, as shown in Fig.~\ref{fig2}(a). At a driving power of $P=0.1\,\textrm{mW}$, $|\alpha|\approx 10^{3}$ and $\beta\approx 40$, consistent with our assumptions for linearization.

In the vicinity of the steady-state amplitudes, the master equation of our optomechanical system has the same form as that in Eq.~(\ref{rho1}) but with $H_{t}$ replaced by a shifted Hamiltonian $H_{\textrm{sh}} = H_{\textrm{eff}}+H_{\textrm{nl}}$. Here
\begin{equation}
H_{\textrm{eff}}=\Delta_{a}a^{\dagger}a+\tilde{\omega}_{m}b^{\dagger}b+\Lambda (b^{2}+b^{\dagger 2})
-G(a+a^{\dagger})(b+b^{\dagger}),\label{Heff}
\end{equation}
only containing linear and bilinear terms with the following coefficients
\begin{align}
& \Delta_a =\delta_a-2g_0\beta,\quad \tilde{\omega}_m =\omega_m+2\Lambda,\nonumber \\
& \Lambda=3\eta(4\beta^2+1),\quad G=g_0|\alpha|; \label{coeff}
\end{align}
at the same time 
\begin{align}
H_{\textrm{nl}}=&- g_0a^{\dagger}a\left(b+b^{\dag}\right) + \frac{1}{2}\eta(b^{\dagger4}+4b^{\dagger3}b+3b^{\dagger2}b^2 \nonumber \\
+& 8\beta b^{\dagger3}+24\beta b^{\dagger2}b+h.c.),\label{Hnl}
\end{align}
composed of all the nonlinear terms generated by the radiation-pressure interaction and the Duffing nonlinearity. The operator $a$ ($b$) here and hereafter is the shifted operator defined relative to the steady-state amplitude $\alpha$ ($\beta$). With $g_0,\, \eta\beta \ll \Lambda, G$, these nonlinear terms in $H_{\rm nl}$ are much weaker than the linear and bilinear terms in $H_{\rm eff}$. After neglecting the nonlinear terms, the master equation becomes 
\begin{equation}
\dot{\rho}=- i[H_{\textrm{eff}},\rho]+\kappa\mathcal{D}[a]\rho+\gamma(\bar{n}_{\textrm{th}}+1)\mathcal{D}[b]\rho +\gamma\bar{n}_{\textrm{th}}\mathcal{D}[b^{\dagger}]\rho,\label{rho}
\end{equation}
governed by the effective Hamiltonian $H_{\rm eff}$ and the damping terms. The third term in $H_{\rm eff}$ describes a parametric-amplification process induced by the Duffing nonlinearity and plays a key role in squeezing generation~\cite{QuantumOptics}. This term can also be viewed as an increase of the spring constant of the mechanical mode. The last term in $H_{\rm eff}$ describes an effective optomechanical coupling that causes cooling and heating of the mechanical mode~\cite{cooling1, cooling2, cooling3, cooling4, cooling5}. 

Parametric-amplification processes induce instability. Applying the Routh-Hurwitz criterion~\cite{stability}, we derive the stability condition for this system:
\begin{equation}
16G^2<(\omega_m+4\Lambda)(4\Delta_a+\kappa^2/\Delta_a),\label{instability}
\end{equation}
for red-detuned driving with $\Delta_a>0$. This condition is satisfied in all relevant parameter regimes in our scheme (see Appendix~\ref{ssec3} for details). Interestingly, at the optimal detuning point for squeezing (see below), this condition can be simplified to be $g_{0}<\sqrt{27\omega_m\eta}$, independent of the driving power $P$. Meanwhile, our parameter regimes are well separated from the bistability threshold for a Duffing oscillator.

\section{Mechanical squeezing~\label{sec3}} 
Apply the squeezing transformation $S(r) = \exp[r(b^{2}-b^{\dagger 2} )/2]$ with squeezing parameter 
\begin{equation}
r=(1/4)\ln(1+4\Lambda/\omega_{m})\label{eq:r}
\end{equation}
to the effective Hamiltonian $H_{\textrm{eff}}$~\cite{nonlinearity2}. Under this transformation, 
\begin{equation}
S^{\dagger}(r)bS(r) = b\cosh(r) - b^{\dag}\sinh(r)
\end{equation}
and $S^{\dagger}(r)aS(r) = a$. The Hamiltonian is hence transformed to be $H_{\textrm{eff}}^{\prime}=S^{\dagger}(r)H_{\textrm{eff}}S(r)$ with
\begin{equation}
H_{\textrm{eff}}^{\prime}=\Delta_{a}a^{\dagger}a+\omega'_{m}b^{\dagger}b-G'(a+a^{\dagger})(b^{\dagger}+b),\label{Heff'}
\end{equation}
where $\omega_{m}^{\prime}=\omega_{m}\sqrt{1+4\Lambda/\omega_{m}}$ is the transformed mechanical frequency and $G^{\prime}=G(1+4\Lambda/\omega_{m})^{-1/4}$ is the transformed optomechanical coupling. In Fig.~\ref{fig2}(b), we plot $\omega_{m}^{\prime}$ and $G^{\prime}$ as functions of the driving power $P$, both of which increase monotonically with $P$. At a driving power of $P=0.1\,\textrm{mW}$, we have $\omega_{m}^{\prime}\approx 3\,\omega_{m}$ and $G^{\prime}\approx\,0.6 G$.

We then apply the squeezing transformation $S(r)$ to the master equation in Eq.~(\ref{rho}) and define the transformed density matrix $\rho^{\prime}=S^{\dagger}(r)\rho S(r)$. It can be shown that $S^{\dagger}(r) \mathcal{D}[a]\rho S(r) = \mathcal{D}[a]\rho^{\prime}$ and
\begin{eqnarray}
&S^{\dagger}(r) \mathcal{D}[b]\rho S(r)  =  \cosh^{2}(r) \mathcal{D}[b] \rho^{\prime}  + \sinh^{2}(r)  \mathcal{D}[b^{\dag}] \rho^{\prime} & \nonumber \\ 
& - \cosh(r)\sinh(r)\left( \mathcal{G}[b] +\mathcal{G}[b^{\dag}]\right) \rho^{\prime}& \label{eq:transb}
\end{eqnarray}
with $\mathcal{G}[o]\rho=o\rho o-(oo\rho+\rho oo)/2$. Similar result can be obtained for the term $S^{\dagger}(r) \mathcal{D}[b^{\dag}]\rho S(r)$. With the condition $\Delta_{a}, \omega_{m}^{\prime} \gg G^{\prime}, \gamma(\bar{n}_{th}+1)$, the $\mathcal{G}[b]\rho^{\prime}$ and $\mathcal{G}[b^{\dag}]\rho^{\prime}$ terms in the above equation are fast oscillating with factors $\sim e^{\pm 2 i \omega_{m}^{\prime} t }$ and can be neglected under the rotating-wave approximation (RWA). The validity of this approximation is manifested in Fig.~\ref{fig5}, where numerical result calculated from the transformed master equation agrees accurately with the result from the original master equation. Hence under the RWA, the transformed master equation for the density matrix $\rho^{\prime}$ has the same form as Eq.~(\ref{rho}) with $H_{\textrm{eff}}$ replaced by $H_{\textrm{eff}}^{\prime}$ and $\bar{n}_{\textrm{th}}$ replaced by 
\begin{equation}
\bar{n}_{\textrm{th}}^{\prime}=\bar{n}_{\textrm{th}}\cosh(2r)+\sinh^{2}(r).\label{nth'}
\end{equation}
Note that the mechanical damping rate is not affected by the squeezing transformation. As the Hamiltonian $H_{\textrm{eff}}^{\prime}$ only contains linear and bilinear couplings between the cavity and the mechanical modes, the transformed master equation for $\rho^{\prime}$ describes a standard cavity cooling process with thermal phonon number $\bar{n}_{\textrm{th}}^{\prime}$~\cite{cooling2, cooling3, cooling4}.

\begin{figure}
\includegraphics[width=\columnwidth,clip]{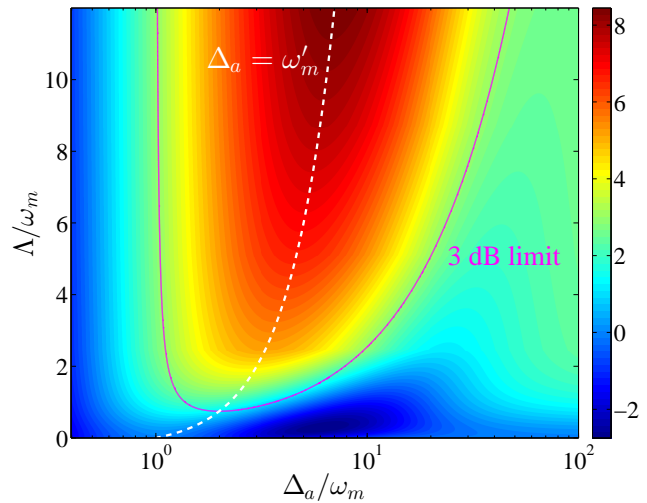}
\caption{(Color online) The squeezing of $X$ (in units of dB) versus $\Delta_a$ and $\Lambda$ at $\bar{n}_{\textrm{th}}=0$. Parameters are the same as in Fig.~\ref{fig2}. The dashed and solid lines correspond to squeezing at the optimal detuning ($\Delta_{a}=\omega_{m}^{\prime}$) and at $3$ dB, respectively.}
\label{fig3}
\end{figure}
The squeezing of the mechanical mode can be calculated by solving the above master equation. The steady-state density matrix $\rho_{\rm ss}^{\prime}$ in the transformed frame can be derived by solving Eq.~(\ref{rho}) numerically. The steady-state average of an arbitrary operator $A$ in the original frame (before the transformation) is $\langle A\rangle=\textrm{Tr}[S^{\dagger}(r)AS(r)\rho_{\rm ss}^{\prime}]$. For the displacement quadrature $X=(b+b^{\dag})/\sqrt{2}$ of the mechanical mode, its steady-state variance can then be derived as
\begin{equation}
\langle\delta X^{2}\rangle_{\rm ss}=\left(\bar{n}_{\textrm{eff}}^{\prime}+\frac{1}{2}\right)e^{-2r},\label{X2}
\end{equation}
where $\bar{n}_{\textrm{eff}}^{\prime}$ is the steady-state phonon number of the transformed system and is determined by the cooling process. Best cooling in the transformed system occurs at the optimal detuning $\Delta_{a}=\omega_{m}^{\prime}$. Hence Eq.~(\ref{X2}) shows that at a given driving power (given $r$ and $\Lambda$), squeezing is strongest at the optimal detuning. This is clearly illustrated by the dashed contour in Fig.~\ref{fig3}. For comparison, we also plot the contour of the $3$ dB limit where $\langle\delta X^{2}\rangle_{\rm ss}=1/4$.

\begin{figure}
\includegraphics[width=\columnwidth,clip]{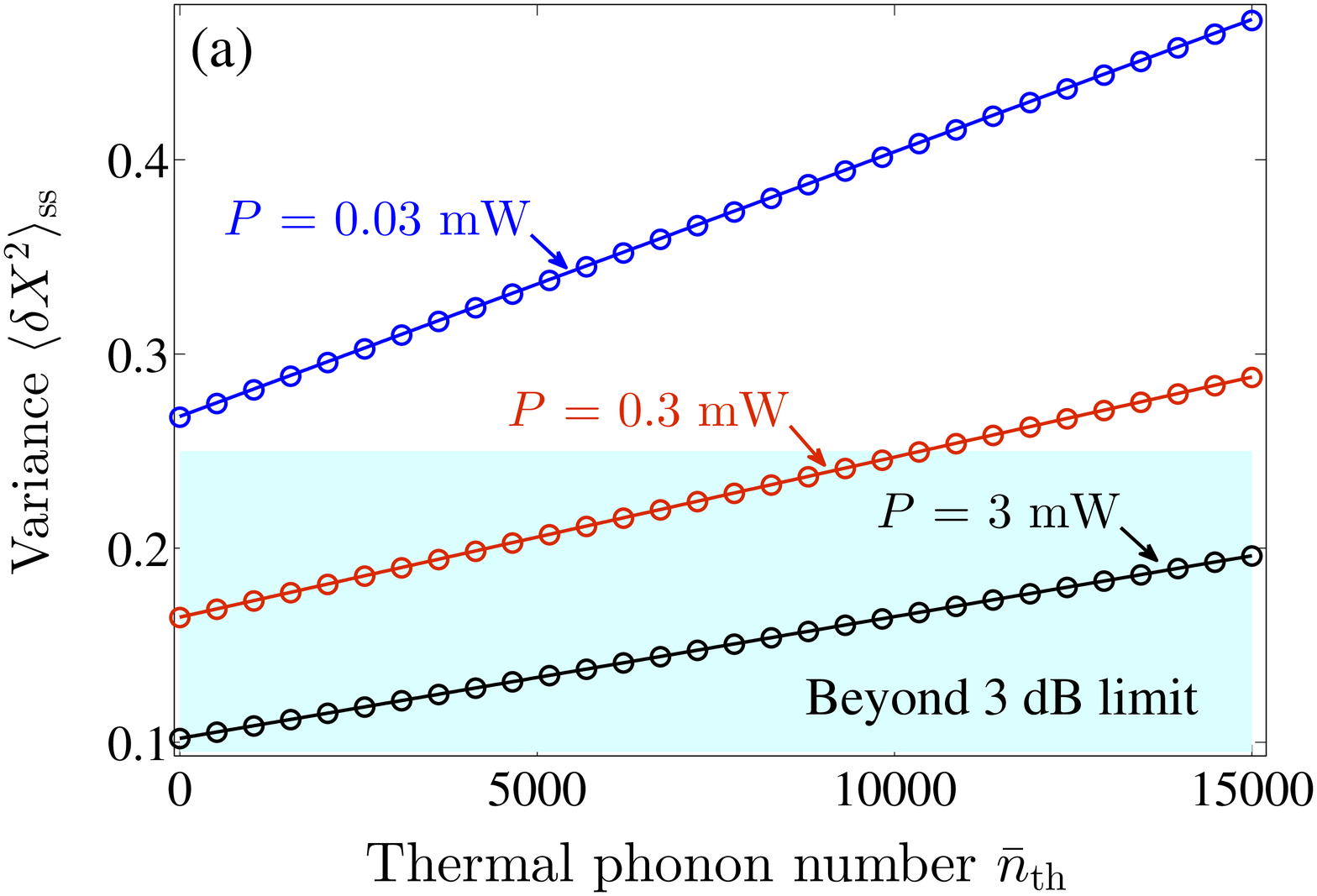}
\includegraphics[width=\columnwidth,clip]{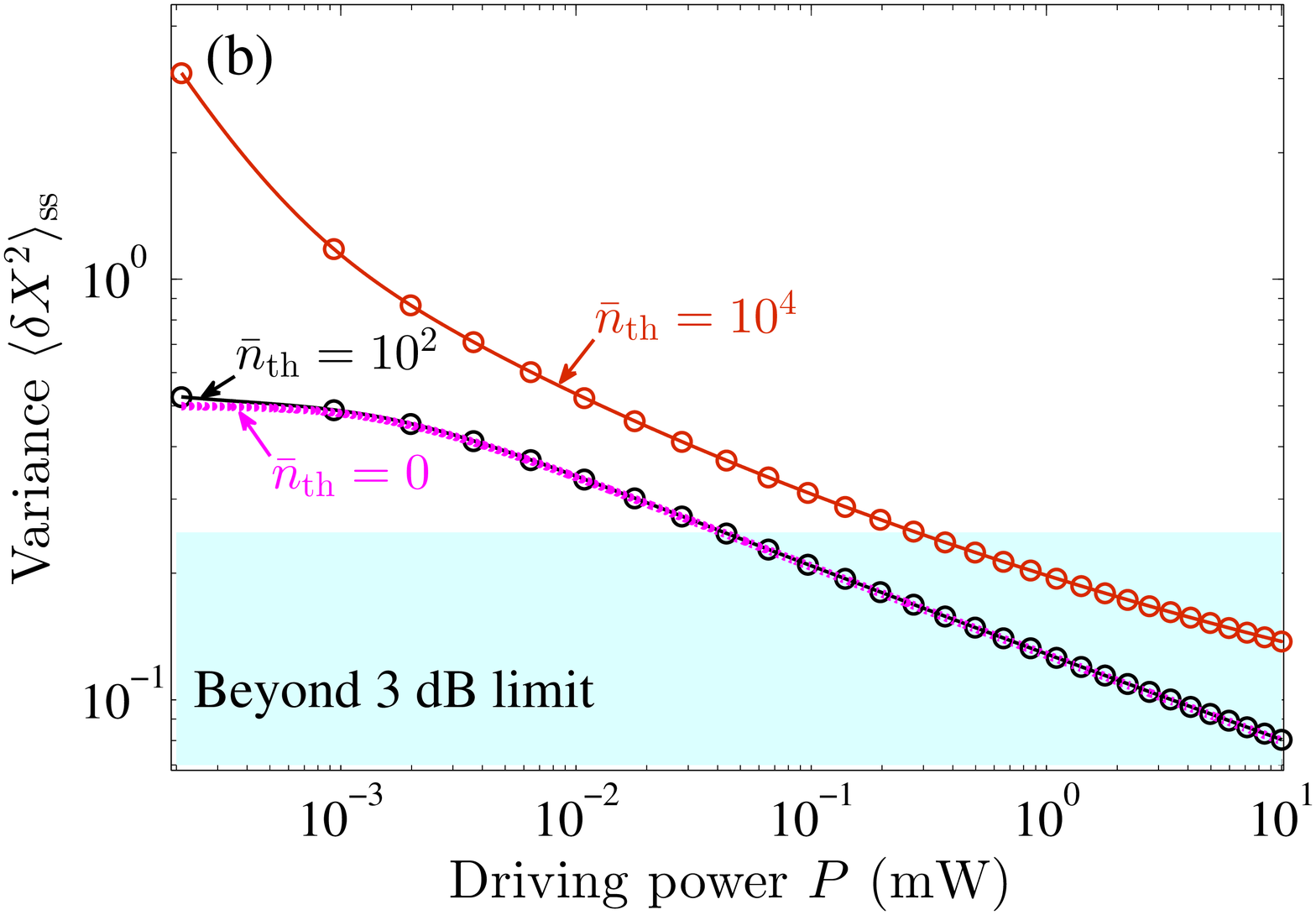}
\caption{(Color online) (a) The variance $\langle\delta X^{2}\rangle_{\rm ss}$ versus $\bar{n}_{\textrm{th}}$ at selected driving powers. (b) The variance $\langle\delta X^{2}\rangle_{\rm ss}$ versus $P$ at selected $\bar{n}_{\textrm{th}}$. All plots are at the optimal detuning. Other parameters are the same as in Fig.~\ref{fig2}. The shadowed blue bottom region corresponds to squeezing beyond the $3$ dB limit. The solid curves (circles) correspond to exact numerical solution (analytical solution in the strong-coupling limit).}
\label{fig4}
\end{figure}
In Fig.~\ref{fig4}(a), we plot $\langle\delta X^{2}\rangle_{\rm ss}$ as a function of the average thermal phonon number $\bar{n}_{\textrm{th}}$. The variance is proportional to $\bar{n}_{\textrm{th}}$ with a slope that decreases with the driving power. This can be explained by Eq.~(\ref{X2}) where the variance increases with the effective phonon number $\bar{n}_{\textrm{eff}}^{\prime}$ which is proportional to $\bar{n}_{\textrm{th}}$. Our result also shows that as the driving power reaches a threshold value, squeezing exceeding $3$ dB can be reached. Even at a high temperature with $\bar{n}_{\textrm{th}}\sim 10^{4}$, strong steady-state squeezing can still be achieved by increasing the driving power. The dependence of the variance on the driving power is shown in Fig.~\ref{fig4}(b), where the mechanical squeezing becomes stronger as the driving power increases.

\section{Analytical Solutions~\label{sec4}}
\subsection{Cooling limit~\label{subsec4a}} 
To better understand the proposed squeezing scheme, we study limiting cases that have analytical solutions. First, consider the limit of $G^{\prime}\ll\kappa\ll\omega_{m}^{\prime}$, where a cooling equation for the mechanical mode can be derived from the master equation in the transformed basis by adiabatically eliminating the cavity mode~\cite{cooling2, cooling3, cooling4}. Let $\mu^{\prime}=\textrm{Tr}_{a}[\rho^{\prime}]$ be the reduced density matrix of the mechanical mode. The cooling equation is
\begin{eqnarray}
\dot{\mu}^{\prime}&=&-i[\omega_{m}^{\prime}b^{\dagger}b,\mu^{\prime}]+[\gamma(\bar{n}_{\textrm{th}}^{\prime}+1)+\Gamma_{-}]\mathcal{D}[b]\mu^{\prime}\nonumber \\
&+&(\gamma\bar{n}_{\textrm{th}}^{\prime}+\Gamma_{+})\mathcal{D}[b^{\dagger}]\mu^{\prime}\label{mu'}
\end{eqnarray}
with the rates
\begin{equation}
\Gamma_{\mp}=\frac{\kappa(G^{\prime})^{2}}{\kappa^{2}/4+(\omega_{m}^{\prime}\mp\Delta_{a})^{2}}.\label{Gammamp} 
\end{equation}
The steady state of Eq.~(\ref{mu'}) is a thermal state with average phonon number 
\begin{equation}
\bar{n}_{\textrm{eff}}^{\prime}=\frac{\gamma\bar{n}_{\textrm{th}}^{\prime}+\Gamma _{+}}{\gamma+\Gamma}, 
\end{equation}
where $\Gamma=\Gamma_{-}-\Gamma_{+}$ is the cooling rate. At the optimal detuning $\Delta_{a}=\omega_{m}^{\prime}$, $\Gamma_{-}=4(G^{\prime})^2/\kappa$, $\Gamma_{+}\approx \kappa(G^{\prime}/2\omega_{m}^{\prime})^{2}$, and strong cooling can be achieved. The density matrix of the mechanical mode in the original basis $\mu=S\mu^{\prime}S^{\dag}$ is hence a squeezed thermal state. The variance of the squeezed mechanical quadrature depends on the squeezing parameter $r$ and the cooling rate $\Gamma$, both of which are determined by the driving power.

\subsection{Strong-coupling limit~\label{subsec4b}} 
Next, we consider the strong-coupling limit with $\kappa\ll G^{\prime}\ll\omega'_{m}$. In this limit, by omitting the counter-rotating terms ($ab+a^{\dag}b^{\dag}$) in the optomechanical coupling, we can derive analytical solution for the squeezing. At the optimal detuning, we obtain
\begin{equation}
\langle\delta X^{2}\rangle_{\rm ss}=\frac{2\gamma\bar{n}_{\textrm{th}}+\gamma+2\Gamma_{\textrm{sc}}e^{-2r}}{4(\gamma+\Gamma_{\textrm{sc}})}\label{X2sc}
\end{equation}
with cooling rate
\begin{equation} 
\Gamma_{\textrm{sc}}=\frac{4(G^{\prime})^2\kappa}{\kappa^{2}+\kappa\gamma+4(G^{\prime})^2}.
\end{equation}
The contribution of the thermal noise in $\langle\delta X^{2}\rangle_{\rm ss}$ is reduced by a factor $~\gamma/2\Gamma_{\textrm{sc}}$ due to the cavity cooling. At zero temperature and with ultra-strong driving (when $e^{-2r}\ll 1$), the squeezing will be ultimately limited by
\begin{equation} 
\langle\delta X^{2}\rangle_{\rm ss}=\frac{\gamma}{\gamma+4\Gamma_{\textrm{sc}}},
\end{equation}
which can be approximated as $\langle\delta X^{2}\rangle_{\rm ss}\approx \gamma/4\kappa$. For a typical optomechanical system with $\gamma\ll\kappa$, this indicates a strong squeezing well below the standard quantum limit. This analytical solution is shown in Fig.~\ref{fig4}. It can be seen that it agrees well with that of exact numerical solution.

\section{Detection of squeezing~\label{sec5}} 
To detect the mechanical squeezing generated in our approach, we consider an ancilla cavity mode $a_s$ (with resonant frequency $\omega_s$) driven by a pump field of amplitude $\Omega_p$ and frequency $\omega_p$, as depicted in Fig.~\ref{fig1}. The frequency separation between the cavity modes $a$ and $a_s$ is much larger the frequency of the mechanical mode, i.e., $|\omega_a-\omega_s|\gg\omega_m$. With the detection circuit included, the total Hamiltonian of this system becomes
\begin{equation}
H_{\rm dec}=H_{t}+\delta_{s} a_s^{\dagger}a_s-g_s a_s^{\dagger}a_s(b^{\dagger}+b)+\Omega_p(a_s^{\dagger}+a_s),\label{detectionH}
\end{equation} 
where $H_{t}$ is given by Eq.~(\ref{H}), $\delta_s=\omega_s-\omega_p$ is the detuning of the ancilla mode $a_s$, and $g_s$ is the strength of the single-photon optomechanical coupling. Under pumping, the ancilla mode reaches a steady-state amplitude $\alpha_{s}$. The effective Hamiltonian is then
\begin{equation}
H^{\textrm{dec}}_{\textrm{eff}}=H_{\rm eff}+\Delta_s a_s^{\dagger}a_s-G_s(a_s+a_s^{\dagger})(b+b^{\dagger}),\label{detectionHeff}
\end{equation}
where $H_{\rm eff}$ is given by Eq.~(\ref{Heff}), $\Delta_s=\delta_s - 2g_s\beta$, and $G_s = g_s\alpha_{s}$. As shown in Ref.~\cite{detection1}, both the position and the momentum quadratures of the mechanical resonator in the original frame (untransformed frame) can be measured by homodyning the output field of the ancilla mode with a local oscillator. Effective detection of the mechanical state requires that $\alpha_{s}\gg 1$ while $G_{s}\ll\kappa_{s}$, where $\kappa_{s}$ is the damping rate of the ancilla cavity mode. Meanwhile, to reduce the detection backaction on the mechanical mode, it requires that $\alpha_{s}\ll\alpha$ when the coupling constants $g_{s}\sim g_{0}$. We choose $P_s\approx0.1\,\mu\textrm{W}$ for an ancilla cavity of $\omega_{s}/2\pi=1000\,\textrm{THz}$, which leads to $\alpha_{s}\approx50$. With these parameters, the output field of the mode $a_{s}$ provides a direct measurement of the quadrature variances of the mechanical resonator. 

A weak force applied to the mechanical resonator can be detected by measuring the output field of the ancilla cavity. The weak impulsive force generates a displacement of the mechanical state in its phase space of the original frame, which can be detected from the output field within a finite time window shorter than the inverse of the cooling rate $\Gamma_{s}=4G_{s}^{2}/\kappa_{s}$. Strong squeezing of the mechanical mode ensures that the detection of this force has a resolution far exceeding the standard quantum limit~\cite{CavesRMP1980,CavesPRD1981}. 

\section{Discussions~\label{sec6}}
\begin{figure}
\includegraphics[width=\columnwidth,clip]{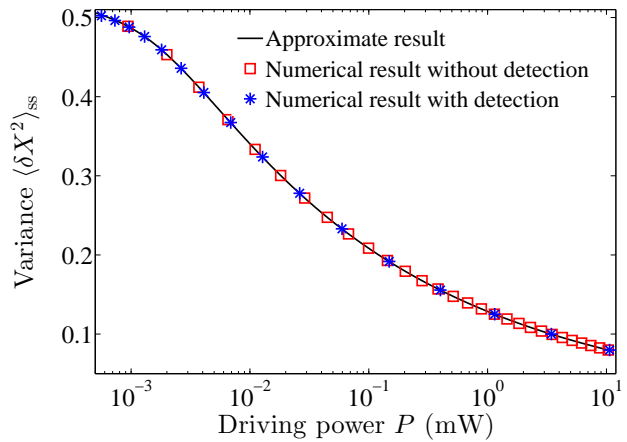}
\caption{The steady-state variance $\langle\delta X^2\rangle_{\rm ss}$ versus the driving power $P$ at $\bar{n}_{\rm th}=10^2$. Solid line: solution under the linearized Hamiltonian $H_{\rm eff}$; squares: with $H_{\rm nl}$ included; asterisks: with detection circuit included. Here $\Delta_a=\omega_m^{\prime}$, $\Delta_s=\omega_m$, and other parameters are the same as in Fig.~\ref{fig2}.}
\label{fig5}
\end{figure}

In the previous sections, we showed that mechanical squeezing robust against thermal noise can be generated under the effective Hamiltonian $H_{\rm eff}$, where the nonlinear Hamiltonian $H_{\rm nl}$ and the backaction of the detection circuit are omitted from the discussion. To evaluate the validity of the linearization procedure, we numerically solve the master equation that includes the nonlinear Hamiltonian and plot the steady-state variance $\langle\delta X^2\rangle_{\rm ss}$ in Fig.~\ref{fig5}. Our results show no distinguishable difference between the solutions with and without the linearization approximation. Similarly, we study the influence of the detection on our squeezing scheme. In Fig.~\ref{fig2}, the steady-state amplitudes $|\alpha|$ and $\beta$ are plotted in the presence of the detection circuit; and in Fig.~\ref{fig5}, the steady-state variance $\langle\delta X^2\rangle_{\rm ss}$ is plotted. Our results show that detection has negligible effect on the mechanical squeezing. 
\section{Conclusions~\label{sec7}} To conclude, we presented a method to generate steady-state mechanical squeezing that is robust against thermal fluctuations. Our approach utilizes mechanical nonlinearity and strong driving on the cavity mode in an optomechanical system. The mechanical squeezing is a consequence of the joint effect of the nonlinearity-induced parametric amplification and cavity cooling. We showed that strong squeezing can be achieved at the optimal detuning where the cavity detuning is in resonance with the transformed mechanical frequency. Analytical solutions in two limiting cases are derived. In a wide range of driving power and thermal phonon number, squeezing well below the standard quantum limit can be achieved. The steady-state squeezing can be detected by measuring the output field of an ancilla cavity mode. 

\begin{acknowledgements}
XYL thanks Prof.~Ying Wu and Prof.~Hui Jing for valuable discussions. XYL and JQL are supported by the JSPS Foreign Postdoctoral Fellowships under No.~P12204 and No.~P12503, respectively. XYL is also supported by NSFC-11374116. LT is supported by the DARPA ORCHID program through the AFOSR, NSF-DMR-0956064, and the NSF-COINS program under No. NSF-EEC-0832819. FN is supported by the RIKEN iTHES Project, MURI Center for Dynamic Magneto-Optics, and Grant-in-Aid for Scientific Research (S).
\end{acknowledgements}

\appendix
\section{Generation of strong Duffing nonlinearity~\label{ssec1}}
In this appendix, we provide detailed discussion on the generation of strong mechanical nonlinearity. Various approaches have been studied to generate strong nonlinearity by coupling the mechanical resonator to an ancilla system~\cite{BlencowePR2004, NoriRMP2013, strong_nonlinear1, strong_nonlinear2}. We focus on the method in Ref.~\cite{strong_nonlinear1}, where the nonlinearity is generated by coupling the mechanical resonator to an ancilla qubit. Consider an ancilla qubit with the Hamiltonian $H_{q}=(\Delta_{q}/2)\sigma_{x}$, which couples to the mechanical mode via an interaction $\lambda_{q} X\sigma_{z}$. This coupling induces an effective Duffing nonlinearity on the mechanical resonator in the form of $H_{m}^{(4)}=6\Delta_{q} (\lambda_{q}/\Delta_{q})^{4}X^{4}$, when the qubit is in an eigenstate of $\sigma_{x}$ and under the condition $\lambda_{q}/\Delta_{q}\ll 1$. With $\Delta_{q}/2\pi=5\,\textrm{GHz}$ and $\lambda_{q}=38\,\textrm{MHz}$, $H_{m}^{(4)}$ gives a nonlinear amplitude $\eta/2\pi\sim 0.2\,\textrm{kHz}$ and $\eta/\omega_m\sim10^{-4}$, close to the parameters we used in our calculation. Note that the second order term induced by the qubit-resonator coupling has been absorbed into the spring constant of the mechanical resonator. For a typical driving power of $P=0.1\,\textrm{mW}$, the dimensionless mechanical displacement in the stationary state is $X\sim 50$. The mechanical mode thus generates a backaction on the qubit in the form of $0.6\,\textrm{GHz}\,\sigma_{x}$, the amplitude of which is much weaker than the detuning of the qubit. Hence, the ancilla qubit can be treated as a passive system that is not affected by the mechanical backaction. 

\section{Squeezing with cubic nonlinearity~\label{ssec2}}
In the main text, we showed that strong mechanical squeezing in the steady state can be generated for a mechanical mode with Duffing nonlinearity. In principle, mechanical nonlinearity in other forms can also be utilized to generate squeezing. In this section, we show that a cubic nonlinearity in the form of $\eta(b+b^{\dagger})^3$ can also be used to generate strong mechanical squeezing.

We start with the linearization procedure for a mechanical mode with cubic nonlinearity. Let us denote the steady-state amplitude of the cavity (mechanical) mode as $\alpha_{c}$ ($\beta_{c}$). We find that these amplitudes satisfy the following nonlinear equations:
\begin{subequations}
\begin{align}
[-i(\delta_c-2g_0\beta_c)-\kappa/2]\alpha_{c}-i\Omega_d=0, \\
12\eta\beta_{\rm c}^2+\omega_m\beta_{\rm c}+3\eta-g^2_0|\alpha_{\rm c}|^2=0,
\end{align}
\end{subequations}
where we have dropped $\gamma$-dependent terms for $\gamma\ll\kappa,\eta$. The quantum master equation in terms of the shifted operators can be written as
\begin{eqnarray}
\dot{\rho}&=&-i[H_{\textrm{sh}}^{c},\rho]+\kappa\mathcal{D}[a]\rho+\gamma\left(\bar{n}_{\rm th}+1\right)\mathcal{D}[b]\rho\nonumber \\
&+&\gamma\bar{n}_{\rm th}\mathcal{D}[b^{\dagger}]\rho,\label{rhoc}
\end{eqnarray}
where the total Hamiltonian has the form 
\begin{equation}
H_{\textrm{sh}}^{c}=H_{\textrm{eff}}^{c}-g_0a^{\dagger}a(b+b^{\dagger})+(3\eta b^{\dagger2}b+\eta b^{\dagger3}+h.c.),\label{Hshc}
\end{equation}
and $H_{\textrm{eff}}^{c}$ is composed of the linear and bilinear terms with 
\begin{eqnarray}
H_{\textrm{eff}}^{c}&=&\Delta_{a}^{c}a^{\dagger}a+\tilde{\omega}^{c}_{m}b^{\dagger}b+\Lambda^{c} \left(b^{2}+b^{\dagger 2}\right)\nonumber \\
&-&G^{c}\left(a+a^{\dagger}\right)\left(b+b^{\dagger}\right).\label{Heffc}
\end{eqnarray}
The parameters in the above equations are
\begin{align}
\Delta_a^{c}&=\delta_a-2g_0\beta_{c},\quad\Lambda^{c}=6\eta\beta,\nonumber \\
\tilde{\omega}_m^{c}&=\omega_m+2\Lambda^{c},\quad G^{c}=g_{0}|\alpha_{c}|.
\end{align}
With $|\alpha_{c}|,\beta_c\gg1$, the nonlinear terms can be neglected and $H_{\textrm{sh}}$ can be approximated by the effective Hamiltonian $H_{\textrm{eff}}^{c}$. 

We want to point out that the Hamiltonian $H_{\textrm{eff}}^{c}$ has exactly the same form as $H_{\textrm{eff}}$ in Eq.~(\ref{Heff}) with its parameters depending on the specific form of the cubic nonlinearity. The squeezing of the mechanical mode can be achieved similarly as in the case of the Duffing nonlinearity.

\section{Stability condition~\label{ssec3}}
In this appendix, we study the stability of our system by applying the Routh-Hurwitz criterion to the equations of motion (the Langevin equations) of this system. Based on the Hamiltonian $H_{\textrm{eff}}$, the equations of motion of this system can be written as
\begin{equation}
\dot{\textbf{R}}\left( t\right) =\textbf{A}\textbf{R}\left( t\right) -\textbf{R}_{in}\left( t\right),\label{dynamicEQ}
\end{equation}
where we introduce the operator vectors $\textbf{R}(t) =(a^{\dagger},a,b^{\dagger},b)^{T}$ for the system operators and $\textbf{R}_{in}(t) =(\sqrt{\kappa}a_{in}^{\dagger},\sqrt{\kappa}a_{in},\sqrt{\gamma}b_{in}^{\dagger},\sqrt{\gamma}b_{in})^{T}$ for the input noise operators, and the matrix $\textbf{A}$ is
\begin{equation}
\textbf{A}=\left(
\begin{array}{cccc}
i\Delta_{a}-\frac{\kappa}{2} & 0 & -iG & -iG \\
0 & -i\Delta_{a}-\frac{\kappa}{2} & iG & iG \\
-iG & -iG & i\tilde{\omega}_{m}-\frac{\gamma}{2} & 2i\Lambda \\
iG & iG & -2i\Lambda & -i\tilde{\omega}_{m}-\frac{\gamma}{2}
\end{array}
\right).\label{A}
\end{equation}
The stability for this system is determined by the eigenvalues of the matrix $\textbf{A}$. If all the eigenvalues of $\textbf{A}$ have negative real parts, then the system is stable.

Based on the fact that the similarity transformation does not change the eigenvalues of a matrix, below we apply a similarity transformation
\begin{equation}
\textbf{V}=\left(
\begin{array}{cccc}
1 & 0 & 0 & 0 \\
0 & 1 & 0 & 0 \\
0 & 0 & \cosh r & -\sinh r \\
0 & 0 & -\sinh r & \cosh r
\end{array}
\right)\label{V}
\end{equation}
with $r=(1/4)\ln(1+4\Lambda/\omega_{m})$ to the matrix $\textbf{A}$. The transformed matrix becomes
\begin{align}
\textbf{A}^{\prime}&=\textbf{V}^{-1}\textbf{A}\textbf{V}\nonumber
\\
&=\left(
\begin{array}{cccc}
i\Delta_{a}-\frac{\kappa}{2} & 0 & -iG^{\prime } & -iG^{\prime } \\
0 & -i\triangle_{a}-\frac{\gamma_{a}}{2} & iG^{\prime } & iG^{\prime } \\
-iG^{\prime } & -iG^{\prime } & i\omega_{m}^{\prime }-\frac{\gamma}{2}
& 0 \\
iG^{\prime } & iG^{\prime } & 0 & -i\omega_{m}^{\prime }-\frac{\gamma}{
2}
\end{array}
\right)\label{A'}
\end{align}
with $G^{\prime}=G(1+4\Lambda/\omega_{m})^{-1/4}$ and $\omega_{m}^{\prime}=\omega_{m}\sqrt{1+4\Lambda/\omega_{m}}$. By calculating the eigenvalues of $\textbf{A}^{\prime}$, we derive the stability condition in the red-detuned regime $\Delta_a>0$ as
\begin{equation}
4\omega_{m}^{\prime}(G^{\prime})^{2}\Delta_{a}-\left[(\omega^{\prime}_{m})^{2}+\frac{\gamma^{2}}{4}\right]\left(\Delta_{a}^{2}+\frac{\kappa^{2}}{4}\right)<0.\label{stability}
\end{equation}
Converting this to the original parameters (before the squeezing transformation), the stability condition can be expressed as 
\begin{equation}
16G^2<(\omega_m+4\Lambda)(4\Delta_a+\kappa^2/\Delta_a),
\end{equation}
after omitting the $\gamma$-dependent term as given in the main text. In order to generate strong squeezing, we are interested in the parameter regime of strong driving with $|\alpha|, \beta\gg1$ and near the optimal detuning point with $\Delta_{a}\sim\omega'_m$. In this regime, Eq.~(\ref{stability}) can be simplified to
\begin{equation}
g_{0}<\sqrt{27\omega_{m}\eta},\label{stabilityapprox}
\end{equation}
which is independent of the driving power. The parameter regime of interest in our scheme always satisfies this condition.

\end{document}